# Can Passenger's Active Head Tilt Decrease The Severity of Carsickness?

- Effect of Head Tilt on Severity of Motion Sickness in a Lateral Acceleration Environment -

Takahiro Wada, Hiroyuki Konno, Satoru Fujisawa and Shun'ichi Doi

Kagawa University, 2217-20 Hayashi-cho, Takamatsu, Kagawa 761-0396, Japan

**Contact information:**

Takahiro Wada, Faculty of Engineering, Kagawa University

2217-20 Hayashi-cho, Takamatsu, Kagawa 761-0396, Japan

Email: wada@eng.kagawa-u.ac.jp, Phone: +81-87-864-2336, Fax: +81-87-864-2369

**Short Version of the Title for Running Heads:** HEAD TILT DECREASES CARSICKNESS

**Keywords:** Motion Sickness, Carsickness, Automobile, Head-Tilt Strategy, Ride Comfort, Vehicle Motion

**Précis:** We investigated the effect of the passenger head-tilt strategy on the severity of carsickness during lateral acceleration situations in automobiles. The results from real car experiments revealed that the active head tilt against centrifugal acceleration has the effect of reducing the severity of motion sickness.

**ABSTRACT**

**Objective**: We investigated the effect of the passenger head-tilt strategy on the severity of carsickness in lateral acceleration situations in automobiles.

**Background**: It is well known that the driver is generally less susceptible to carsickness than are the passengers. However, it is also known that the driver tilts his or her head toward the curve center when negotiating a curve, whereas the passenger's head moves in the opposite direction. Therefore, we hypothesized that the head-tilt strategy has the effect of reducing the severity of carsickness.

**Method**: A passenger car was driven on a quasi-oval track with a pylon slalom while the participant sat in the navigator seat. The experiment was terminated when either the participant felt the initial symptoms of motion sickness or the car finished 20 laps. In the natural head-tilt condition, the participants were instructed to sit naturally, to relax, and not to oppose the lateral acceleration intentionally. In the active head-tilt condition, the participants were asked to tilt their heads against the centrifugal acceleration, thus imitating the driver's head tilt.

**Results**: The number of laps achieved in the active condition was significantly greater than that in the natural condition. In addition, the subjective ratings of motion sickness and symptoms in the active condition were significantly lower than those in the natural condition.

**Conclusion**: We suggest that an active head tilt against centrifugal acceleration reduces the severity of motion sickness.

**Application**: Potential applications of this study include development of a methodology to reduce carsickness.

## INTRODUCTION

Motion sickness such as carsickness and sea sickness decreases the comfort of humans in a vehicle or ship. To reduce motion sickness, it is necessary to clarify its mechanism and to develop a reduction method. Many research studies have been conducted on motion sickness. These studies include experiments involving human exposure to whole-body vibration to examine the susceptibility to motion sickness under vibrations at various frequencies. For example, a vibration analysis with a subjective evaluation revealed some of the characteristics of human susceptibility to motion sickness (Griffin, 1990), and these results were successfully applied to the design of a vehicle control system (Doi et al., 1988; Kajino et al., 2008).

It is known that the riding conditions affect the severity of motion sickness. For example, the automobile driver is generally less susceptible to carsickness than are the passengers. Rolnick and Lubow (1991) investigated the reason that drivers are less susceptible from the viewpoint of controllability of the vehicle. The paper showed that participants who controlled the vehicle were less likely to get motion sickness than participants without control who were "yoked" mechanically. Furthermore, the driver is known to tilt his or her head toward the curve center when negotiating a curve, whereas the passenger's head moves in the opposite direction (Fukuda, 1976; Zikovitz & Harris, 1999; Konno et al., 2010). Some interpretations of the meaning of head-tilt strategies in such situations are possible. One of the interpretations is to obtain visual information of road shape. For example, Zikovitz and Harris (1999) demonstrated that the head tilt has a correlation with visual information of the road curvature. Another interpretation is to reduce the severity of motion sickness. For example, Fujisawa et al. (2009) demonstrated a result that implied that the driver's head tilt toward the centripetal direction has a correlation with the lateral acceleration while curve driving. Golding et al. (2003) investigated the difference in the severity of motion sickness with different head movements in a longitudinal acceleration condition aligned and misaligned with the gravito-inertial force (GIF) in active and passive situations, under the assumption that the GIF has a critical effect on motion sickness. In addition, Golding et al. (1995) investigated the effect of motion direction and body postures such as upright and supine on motion sickness. They found that motion sickness increased when the body axis was aligned with the direction of acceleration. In addition, from the theoretical viewpoint of the mechanism of motion sickness, Wada et al. (2010) previously demonstrated that the estimated motion sickness incidence due to the driver's head motion was less than that due to the passenger's head motion by using the mathematical model of motion sickness proposed by Kamiji et al. (2007). The mathematical model was based on the subjective vertical conflict (SVC) theory, which hypothesizes that motion sickness occurs due to the integration of the discrepancy between the vertical direction of the earth estimated from sensory information and the estimated value calculated by the internal model built in the central nervous system (Bles et al., 1998). The mathematical model by Kamiji et al.(2007) is an expanded version of the mathematical model proposed by Bos and Bles (1998) for 6-DOF head motion including head rotation.

      Based on the findings from the previous literature, we hypothesized that the driver's head movement against the centrifugal direction has an effect in reducing motion sickness. Furthermore, from the viewpoint of the reduction of passengers' motion sickness, we hypothesize that a passenger's active head motion that imitates the driver's head tilt against centrifugal acceleration has the effect of decreasing the severity of motion sickness. The purpose of the present paper is to examine this hypothesis. Thus, this paper investigates the effect of the

passenger's active head tilt on the severity of motion sickness in lateral acceleration situations by conducting experiments with a real passenger car. It is expected that the results can provide a basic knowledge for establishing vehicle design and control to minimize motion sickness or to increase comfort.

# METHOD

**Design**

The experimental design was mainly based on the work of Golding et al. (2003). Ten participants were exposed to an acceleration stimulus as passengers seated in the automobile navigator seat. Each participant made head motions in either the abovementioned natural or active condition. In the natural condition, the participants were instructed to sit naturally, to relax, and not to oppose the lateral acceleration intentionally. An experimenter sat in the rear seat and asked the participant to be more relaxed ("natural") if the participant tilted his/her head against the centrifugal acceleration as much as the driver did. In the active condition, the participants were asked to tilt their head against the centrifugal acceleration, thus imitating the driver's head tilt. Figure 1 illustrates the typical head postures in the natural and the active conditions. However, the participants were not instructed on the amplitude or the timing of the head tilt. The head-tilt condition was treated as a within-subject factor. Each participant experienced two head-tilt levels on two different days at least three days apart in a crossover design. The order of the conditions was counter-balanced to remove the order effect. In addition, each participant attended the experiments at the same time on the two days if possible, but the experiment days were determined by the participants' schedules. The experiments used two drivers. Each participant was assigned to the same driver on both experiment days.

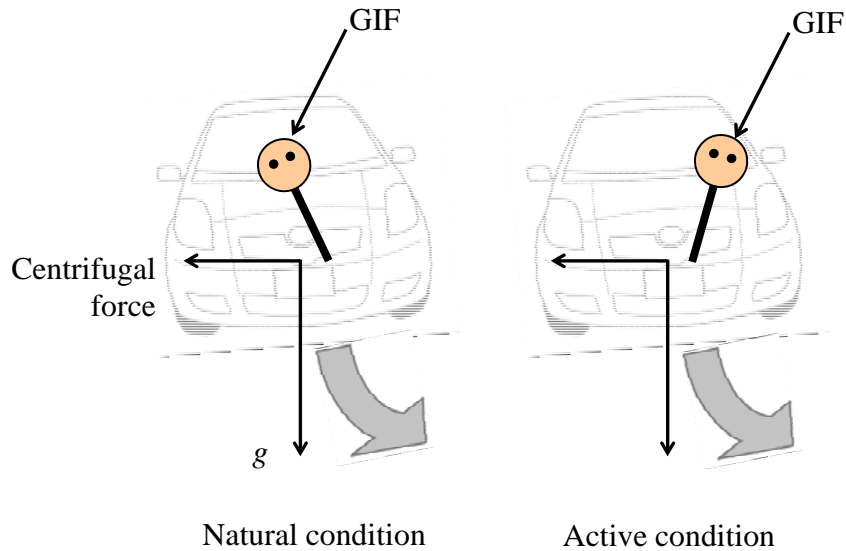

*Figure 1*. Typical head postures in the natural and active conditions

**Participants**

Ten healthy individuals, nine males and one female, with a mean age of 21.5 yr (SD 1.0 yr) gave informed consent to participate in the experiments. The participants were told that they could become motion sick and vomit because of the experiments. The explanation given to the participants also stated that they could stop the experiment at any time and for any reason. One participant had already experienced carsickness on the way to the experimental course. The participant confessed so after the experiments. Thus, the results for that participant were not analyzed. The participants were paid approximately $110 for their participation. Before the experiments, the motion sickness susceptibility of each participant was tested using a revised version of the Motion Sickness Susceptibility Questionnaire (MSSQ: Golding, 1998). The mean percentile score was 54.2% (SD 26.3%). This illustrates that the susceptibility to motion sickness of the participants had a wide distribution. For female individuals, it has been found that the

menstrual cycle affects their susceptibility to motion sickness (Grunfeld, 1998). We treated the data for the female participant the same as that of the males because the female participant was neither in the most susceptible time (around menstruation, at days 3 to 7 and days 25 to 27) nor in the least susceptible time (around ovulation, days 11 to 15).

**Method of evaluating motion sickness**

Subjective evaluations and symptom score tests were carried out to evaluate the severity of motion sickness.

For the subjective evaluations, the subjective sickness rating method used in the research studies of Golding et al. (1995, 2003) was employed. In this method, the severity of motion sickness was rated by a Likert-type scale on the following six levels: 1 = no symptoms, 2 = initial symptoms but no nausea, 3 = mild nausea, 4 = moderate nausea, 5 = severe nausea, and 6 = vomiting. In the present paper, the method is referred to as the *six-level sickness rating*. This method is suitable for the real-time rating of motion sickness because the ratings can be determined easily and in a very short time. At the end of every lap in the experiments, the experimenter sitting in the rear seat asked the participants to give his or her current state according to the six-level sickness rating. In addition, the participants were also asked to declare if they felt any symptom of motion sickness at any time during a lap. A rating indicating motion sickness triggered the termination of that driving trial.

The symptoms of motion sickness were quantified by a motion-sickness symptom score test (Golding et al., 1995) administered approximately 5 min and 10 min after the termination of the driving test. The score test quantifies the symptoms of motion sickness, rated subjectively by the participants using four levels of 0 = none, 1 = mild, 2 = moderate, and 3 = severe for each symptom of dizziness, body warmth, headache, sweating, stomach awareness, increased salivation, nausea, pallor (evaluated by the experimenter), and any additional symptoms. The *total symptom score* was calculated by summing the ratings to quantify the severity of motion sickness (Golding et al., 1995).

**Apparatus**

A small passenger car with a 2.37 m wheel base and a 1000 cc engine was used for the driving experiments. An MTi-G sensor (Xsense Technologies) was attached to a flat place close to the shift lever of the automatic transmission to measure the 3-DOF acceleration and the 3-DOF orientation of the vehicle (Fig. 2). An MTx sensor (Xsense Technologies) was attached to the cap worn by the participants to measure the 3-DOF acceleration and the 3-DOF orientation of the head (Fig. 2). Both sensors were connected to a laptop PC in the rear seat of the vehicle to synchronize the sensor data. The sampling time for the two sensors was 10 ms.

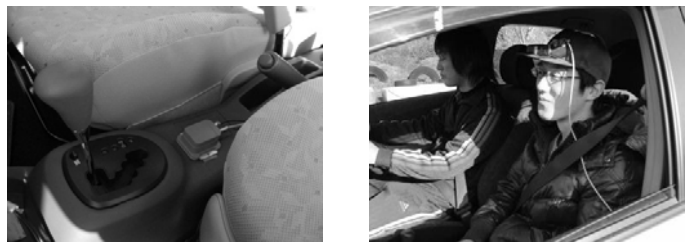

*Figure 2*. Motion sensors attached to the vehicle (left)
and the participant's head (right)

**Procedure**

As stated above, the participants gave written informed consent before the experiments. Each participant was seated in the navigator seat of a small passenger car in a normal seating position with a safety belt and then was exposed to the acceleration applied by the driver. The experimental course was a quasi-oval track with straight parts approximately 100 m in length and curved parts of 8 m and 10 m radii (Fig. 3). Five pylons with 15 m gaps were located in each straight segment. The driver drove the track continuously at approximately 30 km/h through the pylon slalom (i.e., zigzagging to the left and right of the pylons) in the straight segment and at approximately 15 km/h in the curved segments. The track distance was 260 m. The resultant mean lap time and mean total exposure time for 20 laps were 45 s and 14 min 46 sec, respectively. The drivers were aware the head-tilt conditions. The mean frequency of lateral oscillation frequencies in the slalom part and the root mean square (RMS) vehicle lateral accelerations in the head-tilt conditions were compared to show the uniformity of the stimulus. The means and standard deviations of the resultant lateral oscillation frequencies in the natural and active conditions were 0.238 Hz (SD 0.02) and 0.237 Hz (SD 0.02), respectively. The mean frequencies were in the frequency range (around 0.2 Hz) that most provoke the sickness (Golding et al., 2001). No significant differences were found in the lateral oscillation frequencies by the repeated-measures ANOVA ($F(1,15) = 0.082$, $p = 0.778$). The RMS vehicle lateral accelerations in the natural and active conditions were 2.40 m/s$^2$ (SD 0.18) and 2.42 m/s$^2$ (SD 0.19), respectively. No significant differences were found in the RMS vehicle accelerations by the repeated-measures ANOVA ($F(1,7) = 0.024$, $p = 0.881$). Note that the vehicle acceleration data were analyzed for only eight participants due to the misrecording of the vehicle data of one participant in addition to the participant who was not analyzed due to his poor health. At the end of each lap, the experimenter asked the participants to indicate their severity of motion sickness according to the six-level sickness rating used in the research studies of Golding et al. (1995, 2003). The participants were also asked to declare if the six-level sickness rating reached 2 or more at any time during a lap. When level 2 or higher was experienced, the driving trial was terminated at the end of the next lap. Note that the participants were told again at that time that they could stop immediately without this additional lap, but no participant rejected the additional lap. Each driving trial was terminated after 20 laps, the maximum number of laps even if the subjective rating did not reach 2 or more. The time of termination is called the driving endpoint.

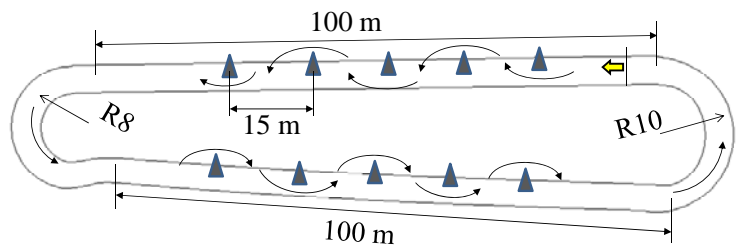

*Figure 3.* Test track

## RESULTS

**Resultant head movement**

Figure 4 presents an example of the time history of the vehicle lateral acceleration and the head tilt. The head-roll angle, defined as the orientation of the head in the frontal plane, was used for evaluation of the head-tilt behavior of the participants. In the natural condition, passive head movement toward the centrifugal direction was found with a time delay. A small head roll without synchronization with the lateral acceleration was also found in the natural condition (no figure). In the active condition, the head-roll angle was found to synchronize with the vehicle lateral acceleration.

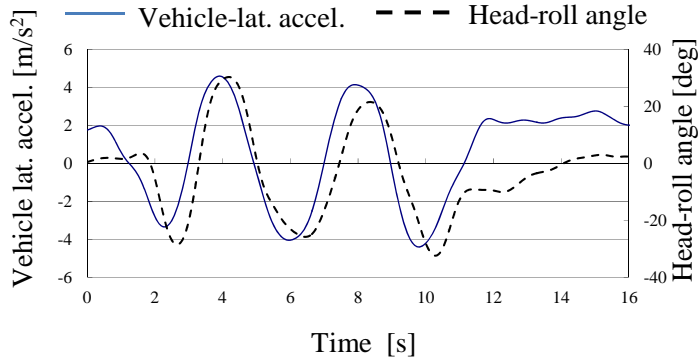

(a)     Natural condition

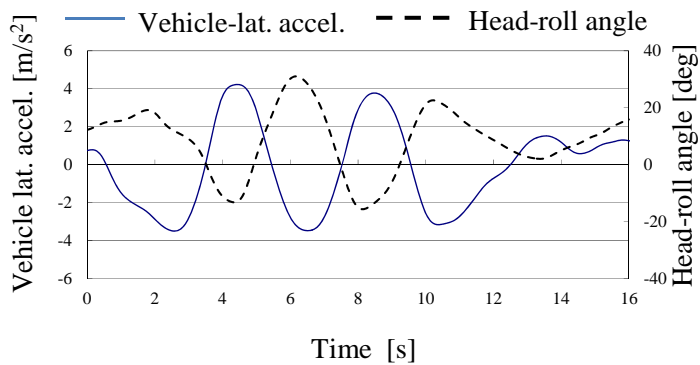

(b)     Active condition

*Figure 4.* Vehicle lateral acceleration and head-roll angle

The correlation coefficients between the vehicle lateral acceleration and the head-roll angle throughout the experiment were analyzed. The means and standard deviations of the correlation coefficients for all participants in the natural and the active conditions were 0.43 (SD 0.30) and -0.71 (SD 0.10), respectively. The positive correlation coefficient represents that the head-roll motion synchronized well with the vehicle lateral acceleration in the centrifugal direction. Thus, a high negative correlation was found in the active condition while a lower positive correlation was found in the natural condition. A Wilcoxon signed-rank test revealed the significance between the head-tilt conditions ($z = -2.52$, $n = 8$, ties = 0, $p = 0.0117$, two-tailed): the head-roll angle in the active condition was greater in the centripetal direction whereas that in the natural condition was made in the opposite direction at lower synchronization with the vehicle motion. This result is consistent with the tendency of the passenger's and driver's head-roll angles in Konno et al. (2010).

**Laps to driving endpoint**

The means and standard deviations for the number of laps of the driving endpoint for all participants in the natural and the active conditions were 13.7 laps (SD 5.6) and 17.9 laps (SD 3.1), respectively. The Wilcoxon signed-rank test was employed to analyze the differences. Some participants recorded 20 laps in both head-tilt conditions. In order to break ties, the number of laps was weighted using the total symptom score recorded 5 min after the experiments by adopting the method of Golding et al. (1995). The Wilcoxon signed-rank test was used for statistical test because of the abnormal distribution of the data, which came from the termination of driving at 20 laps. The Wilcoxon test (with the weighting) revealed significant differences in the number of laps based on the head-tilt condition ($z = -2.53$, $n = 9$, *ties* = 1, $p = 0.011$, two-tailed). The active condition had a larger number of laps to the driving endpoint. The data of the male participants was analyzed by excluding the female participant's data with the Wilcoxon signed-rank tests (with the weighting). The results also revealed significant differences in the number of laps based on the head-tilt condition ($z = -2.38$, $n = 8$, ties = 1, $p = 0.018$, two-tailed).

**Six-level sickness rating at driving endpoint**

TABLE 1 shows the number of participants experiencing each sickness rating level at the driving endpoint. No participant experienced levels 4–6 in the experiments. Five participants experienced level 3 in the natural condition, and six participants experienced no symptoms in the active condition. The data of the male participants was analyzed by excluding the female participant's data with the Wilcoxon signed-rank tests. The results also revealed significant differences in the sickness rating levels between the head-tilt condition ($z = -2.07$, $n = 8$, ties = 3, $p = 0.038$, two-tailed).

*TABLE 1*. Number of Participants Experiencing Each Sickness Rating Level at the Driving Endpoint

| Six-level Sickness Rating | Natural | Active |
|---|---|---|
| 1 No Symptom | 3 | 6 |
| 2 Initial Symptom | 1 | 3 |
| 3 Mild Nausea | 5 | 0 |
| 4 Moderate Nausea | 0 | 0 |

The Wilcoxon signed-rank test revealed a significant difference in the sickness rating levels between the head-tilt conditions ($z = -2.27$, $n = 9$, ties = 3, $p = 0.023$, two-tailed). The six-level sickness rating in the active condition was lower than that in the natural condition.

**Total symptom score**

Figure 5 illustrates the transition of the total symptom scores for both head-tilt conditions. The Wilcoxon signed-rank test revealed a significant difference in the symptom scores between the head-tilt conditions 5 min after driving termination ($z = -2.56$, $n = 9$, ties = 1, $p = 0.011$, two-tailed). The scores in the active condition were lower than those in the natural condition. A significant difference also appeared between the head-tilt conditions 10 min after driving termination ($z = -2.12$, $n = 9$, ties = 4, $p = 0.034$, two-tailed). The data of the male participants was analyzed by excluding the female participant's data with the Wilcoxon signed-rank tests. The results also revealed significant differences in the symptom scores between the head-tilt conditions 5 min after driving termination ($z = -2.39$, $n = 8$, ties = 1, $p = 0.017$, two-tailed) and 10 min after driving termination ($z = -2.12$, $n = 8$, ties = 3, $p = 0.034$, two-tailed).

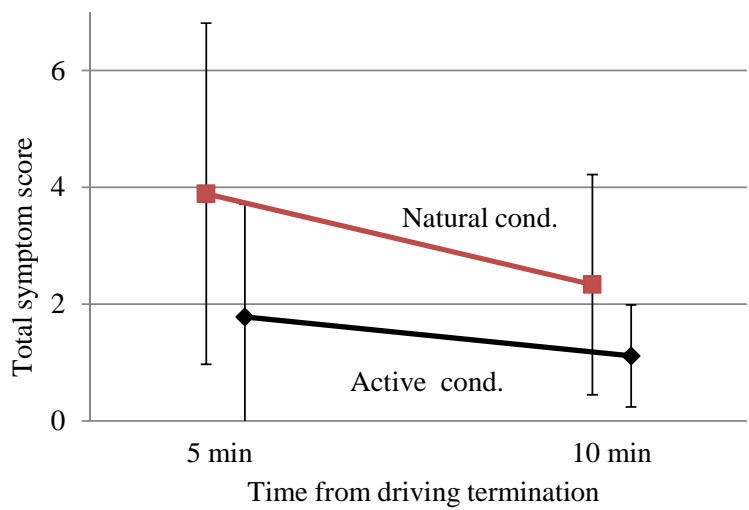

*Figure 5.* Mean total symptom scores

## DISCUSSION

The increase of the number of laps to the driving endpoint and the reduction of the six-level sickness rating and the total symptom scores indicate the effect of the active head-tilt condition in reducing motion sickness. These results suggest that the carsickness of passengers can be reduced if they tilt their head against the centrifugal direction, thus imitating the driver's head tilt.

This finding agrees with the results from Golding et al. (2003), in which the severity of motion sickness decreased when the head was aligned with the GIF in a longitudinal linear acceleration environment in the active head movement of the participant. The contribution of the present paper is to demonstrate the effect of head tilt on reducing the severity of motion sickness in a real automotive environment with visual information of upcoming road shape under lateral acceleration. In addition, the amplitude of the head movement in the present study was not instructed and the head-tilt angle was smaller than that aligned with the GIF. The findings of the present paper are consistent with the hypothesis presented by Fukuda (1976) based on observation of the head movements of bus drivers and their passengers.

Rolnick and Lubow (1991) reviewed in detail the literature relating to immunity from motion sickness in drivers from the viewpoint of factors of head movement, visual information, perceived control, activity, predictability, and controllability. Among these factors, visual information, perceived control, predictability, and controllability were equivalent through the experiments in the present study. The visual information was not very different between the two head-tilt conditions because participants sat in the navigator seat and observed the road scene ahead. A perceived control or sense of control is a subjective psychological state by which the person can determine his or her behavior. The factor of perceived control was also equivalent in both head-tilt conditions because the participants made their head motions by themselves and had been informed that they could stop the experiment at any time. The predictability factor was also equivalent because the participants watched the same scene and rode along a predetermined test track. With regard to the controllability factor, Rolnick and Lubow (1991) showed that participants who felt themselves in control were less likely to get motion sickness. In their paper, a passive participant with no controllability of vehicle motion was also prevented from moving his or her head voluntarily by being "yoked" to the other (active) passenger's head. The present paper reveals that motion sickness can be reduced by a voluntary head tilt toward the GIF direction without any controllability of vehicle motion under the condition where visual information of upcoming road curve is available. Activity is also a factor related to motion sickness, as Wendt (1951) postulated, whereas Rolnick and Lubow (1991) pointed out that very few studies directly indicated the effect of this factor, and one paper demonstrated contradictory results. In the present study, activity is relatively greater in the active head-tilt condition than in the natural condition, since in the active condition the participant needs to control his or her head according to the lateral acceleration changes. This is a limitation of the present study. It is thought, however, that the difference was not very large because appropriate postural control, including head control, was needed even in the natural head-tilt condition. Another possible interpretation of the reduction effect of the motion sickness is difference of active and passive head movements in the head tilt condition. According to reafference principle (von Holst, 1954), adaptation is promoted in the case of the active movement by comparing the efference copy of the motor command and the reafference from the effector. Held & Bossom(1961) showed the prism adaptation was significantly faster in the self-produced motion or active condition. Namely, it can be understood that the rearrangement in the sensory rearrangement theory (Reason & Brand, 1975) was accelerated in the active movements. The active and the natural conditions in the present paper are regarded as the active and the passive movements of the head, respectively. Therefore, the reduction effect of the motion sickness can be partially interpreted as the effect of the active movement. It should be, however, noted again that Golding et al. (2003) showed that the motion sickness by the head motion aligned with the GIF direction was significantly smaller than that with the misaligned head motion even though both head movements are actively produced. This result can be understood that the head movement itself has the effect to reduce motion sickness. To clarify the pure effect of the head tilt of the passenger, experiments with the different head condition including misaligned with the GIF direction should be conducted as a future research study.

Bles et al. (1998) proposed an SVC theory as the mechanism of motion sickness and postulated that motion sickness is caused by accumulation of the discrepancy between the vertical direction sensed by the sensory organs such as the eyes, the vestibular system, and nonvestibular proprioceptors and the subjective vertical direction estimated from the internal model. Wada et al. (2010) found that a head tilt toward the curve center has the effect of reducing the motion sickness incidence by using the mathematical model based on the SVC theory derived by Kamiji et al. (2007), which is a 6-DOF version expanded from the original 1-DOF version by Bos et al. (1998). The result of the SVC model calculation, in which the head movement toward curve center reduces motion sickness incidence agreed with the experimental results of the present paper. In the calculation of both the active and the passive condition, only head movements are changed even though the 6DOF-SVC model includes the parameters reflecting efference copy. Thus, it is understood that a head tilt close to the GIF direction results in a reduction of the conflict with the subjective vertical direction. Thus, the results presented in this paper can be understood as the reduction of motion sickness in the sense of SVC theory.

## CONCLUSION

An experiment was conducted to examine the hypothesis that a passenger's head tilt, imitating a driver's active head-tilt strategy against centrifugal acceleration in an automobile under lateral acceleration, reduces motion sickness. The results showed that the mean number of laps before any symptoms was significantly greater in the active head-tilt condition. In addition, the six-level sickness rating at the driving endpoint in the active head condition, as well as the total symptom score in the active head condition, was significantly reduced. These results strongly indicate that a passenger's active head tilt toward the centripetal direction has the effect of reducing the severity of motion sickness under the lateral acceleration of an automobile.

As a topic for future study, it is important to investigate whether the conclusions are derived for females and a wide range of age groups because the present paper investigated only young people with one female. In addition, the effects of various head motions on motion sickness, including the timing and magnitude of any head tilt, should be investigated in detail. These investigations would lead to basic knowledge for developing a methodology to reduce motion sickness. Experiments with the fixed-head condition should be conducted to investigate the pure effect of the head movement. Furthermore, the comparison between the eyes open and eyes closed conditions would be beneficial to further investigate the basic mechanism of carsickness.

# KEY POINTS

- In comparison with the natural head-tilt condition, the active head tilt of a passenger in a lateral acceleration environment in an automobile significantly increased the number of laps before experiencing the initial symptom of motion sickness.
- The active head-tilt condition significantly reduced the six-level sickness rating and the total

symptom score after the driving trials.

- These results strongly indicate that a passenger's active head tilt toward the centripetal direction has the effect of reducing the severity of motion sickness under lateral acceleration of an automobile.

Biographies

Takahiro Wada is a full professor at the College of Information Science and Engineering, Ritsumeikan University. He received his Ph.D. in Robotics at Ritsumeikan University, Japan, in 1999. He was a research associate at the Department of Robotics, Ritsumeikan University in 1999. He had been an assistant professor at the Faculty of Engineering of Kagawa University from 2000 to 2003 and an associate professor from 2003 to 2012. From 2012, he rejoined Ritsumeikan University as a full professor. He spent a half year in 2006 and 2007 at The University of Michigan Transportation Research Institute as a Visiting Researcher. His current research interests include human-machine systems, human modeling, and driver assistance system for traffic safety.

Hiroyuki Konno is a master course student at the Graduate School of Engineering, Kagawa University. He received his B.S. degree in Intelligent Mechanical Systems at Kagawa University, Japan, in 2011. His current research interests include prevention of carsickness and increase of ride comfort.

Email: twada@fc.ritsumei.ac.jp

Satoru Fujisawa is an engineer at Aisin Seiki Co., Ltd. He received his M.S. degree in Intelligent Mechanical Systems at Kagawa University, Japan, in 2011. He worked for prevention of the ride comfort of automobiles when he was a master course student at Kagawa University.

Shun'ichi Doi is a professor at the Faculty of Engineering at Kagawa University. He received his MS degree and Ph.D. degree in Mechanical Engineering from Nagoya Institute of Technology in 1972 and 1984, respectively. From 1972, he joined Toyota Central R&D Labs, Inc. He joined Kagawa University since 2004. His current research interests include vehicle dynamics and active safety technology.